\documentclass{PoS}
\newcommand{\KK}{{$KK$}}
\newcommand{\KKMC}{\KK MC}
\title{HERWIRI2: CEEX Electroweak Corrections in a Hadronic MC}

\ShortTitle{HERWIRI2: CEEX Electroweak Corrections in a Hadronic MC}

\author{\speaker{Scott Yost}%
         \thanks{This work and its presentation were supported in part by 
D.o.E. grant DE-PS02-09ER09-01 and grants from The Citadel Foundation.}\\
        The Citadel\\
        Charleston, SC 29409, USA\\
        E-mail: \email{scott.yost@citadel.edu}}

\author{Valerie Halyo
	\thanks{Work supported in part by D.o.E grant DE-FG02-91ER40671.}\\
        Princeton University\\
        Princeton, NJ 08544, USA\\
        E-mail: \email{valerieh@princeton.edu}}

\author{Miroslav Hejna\\
        Princeton University\\
        Princeton, NJ 08544, USA\\
        E-mail: \email{mhejna@princeton.edu}}

\author{B.F.L.\ Ward%
	\thanks{Work supported in part by D.o.E. grant DE-FG02-09ER41600.}\\
        Baylor University
        Waco, TX 76798, USA\\
        E-mail: \email{bfl\_ward@baylor.edu}}

\abstract{Reaching the 1\% precision level for $W$ and $Z$ production 
calculations for the LHC will require a mixture of higher order QCD and 
electroweak corrections. As a first step toward implementing the combined QED 
$\otimes$ QCD exponentiation proposed in previous work, we have implemented the 
${\cal O}(\alpha)$ electroweak corrections and YFS exponentiation structure of 
the \KK\ Monte Carlo in HERWIG. We discuss the current status of this program 
and sketch the further developments needed to reach the desired precision level.
}

\FullConference{10th International Symposium on Radiative Corrections 
	(Applications of Quantum Field Theory to Phenomenology)\\
                 September 26-30, 2011\\
                 Mamallapuram, India\hfill
                 \hbox{\tt BU-HEPP-11-06}}

\begin{document}

Vector boson production is one of the most important Standard Model processes
observed at the LHC, and electroweak radiative corrections will be needed for 
analysis at the percent level.  Previous studies \cite{EWsys} 
by some of the authors have
found that electroweak corrections alone can exceed 1\% for some cuts of 
interest. These studies were based on HORACE, \cite{horace}
 which provides state-of-the-art
${\cal O}(\alpha)$ radiative corrections with a final-state photon shower, and
PHOTOS, \cite{photos} which adds final state photonic radiation.
Other programs developed for implementing electroweak corrections for hadronic 
collisions include WINHAC \cite{winhac}
and ZINHAC\cite{zinhac}. Initial state photonic radiation has been considered 
only in certain MRST parton distribution functions. \cite{MRSTQED} However, none
of the most recent PDFs include QED effects. 

For electron-positron colliders, precision electroweak corrections have 
been implemented in the program \KKMC\ \cite{KKMC} 
using YFS \cite{YFS} exponentiated multiple-photon 
radiation for both the initial and final state, together with 
${\cal O}(\alpha)$ electroweak corrections in the DIZET \cite{DIZET} 
package developed for ZFITTER\cite{ZFITTER}. 
The DIZET corrections can be applied to any parton-level process,
so there is no obstacle to extending them to hadronic collisions. YFS
exponentiation provides the basis for an efficient representation of 
multi-photon phase space, which can also be implemented for hadronic 
initial and final states. Some of the authors have proposed using a
non-abelian extension of YFS exponentiation as a basis for an integrated 
event generator implementing both multi-photon and multi-gluon corrections 
in a unified framework, called QCD$\otimes$QED exponentiation. \cite{QCDxQED}  

The collection of programs implementing QCD$\otimes$QED exponentiation has
been called HERWIRI, for High Energy Radiation with Infra-Red Improvements,
with a version number distinguishing the class of 
corrections included.  The name acknowledges that the initial versions build
upon the HERWIG \cite{HERWIG} parton shower generator.  
The first to be released, HERWIRI1, \cite{HERWIRI1} implemented IR-improved 
splitting kernels \cite{kernels} obtained using the QCD analog of YFS 
exponentiation.  This program is publicly
available, and tests are in progress.  The IR-improved kernels have also 
been implemented \cite{HERWIRI1NLO} in MC$@$NLO \cite{MCNLO} and 
POWHEG\cite{POWHEG}.   The structure of HERWIRI is not
tied to a particular shower, and our ultimate goal is a complete 
shower generator based entirely on QCD$\otimes$QED exponentiation
 with exact ${\cal O}(\alpha_s^2, \alpha_s \alpha, \alpha^2)$ 
residuals. \cite{QTS3-RADCOR05}

The second version, HERWIRI2, 
implements the electroweak radiative corrections of \KKMC\ in a hadronic 
shower generator. This note describes a version of HERWIRI2 which is presently 
nearing completion.\footnote{HERWIRI2 does not incorporate HERWIRI1, although
the two programs can be used in combination.}  HERWIRI2 is motivated by the 
successful application of YFS exponentiation in BHLUMI\cite{bhlumi},
BHWIDE\cite{bhwide}, KoralZ\cite{koralz}, KoralW\cite{koralw},
YFSWW3\cite{yfsww3}, \KKMC, and related programs for LEP physics.
All of these programs benefit from a very efficient
representation of $N$-photon phase space, with complete control over the 
soft and collinear singularities for an arbitrary number of photons. 
Real and virtual IR singularities cancel exactly to all orders. The non-abelian
extension to QCD$\otimes$ QED should have similar advantages. 

While based on HERWIG, HERWIRI2 is largely independent of the underlying shower.
HERWIG generates the parton momenta and shower, and HERWIRI2 passes the 
generated hard process momenta to \KKMC\ routines to add photons and 
electroweak corrections.  Although \KKMC\ was developed for $e^+ e^-$ 
collisions, it was designed to be extended to more general processes, so the
ability to select quarks as the incoming state already exists in all but the
lowest-level generation routines.  

\KKMC\ is a precision generator for 
$e^+ e^- \rightarrow f{\overline f}
+ n\gamma,$ $f = \mu,\tau, d, u, s, c, b $ for CMS energies from $2m_\tau$ to
1 TeV.  The precision tag for LEP2 was 0.2\%.
 ISR and FSR $\gamma$ emission is calculated up to ${\cal O}(\alpha^2)$,
including interference.  The MC structure is based on YFS exponentiation, 
including residuals calculated perturbatively to the relevant orders in 
$\alpha^k L^l$.  ($L = \ln(s/m_e^2)$).  Exact collinear 
bremsstrahlung is implemented for up to three photons. 
Electroweak corrections \cite{ZF1,ZF2,ZF3}
are included via DIZET 6.21\cite{DIZET}, and beamstrahlung can be modeled 
over a wide range of energies via a built-in or user-defined distribution.

There are two modes of operation: 
exclusive exponentiation (EEX) and coherent exclusive exponentiation (CEEX).
EEX applies YFS exponentiation to differential cross-sections, while CEEX 
applies it at the amplitude level.  The CEEX mode is written in a manner that
is most readily extended to quark scattering, so it is taken as the basis for
HERWIRI2.  The orders of residuals included in CEEX mode are
$\alpha, \alpha L, \alpha^2 L^2,$ and $ \alpha^2 L.$  

CEEX was introduced for
pragmatic reasons, because the traditional (EEX) exponentiation of spin-summed
differential cross sections suffered from a proliferation of interference terms
in processes with multiple diagrams, limiting its utility in reaching the desired
0.2\% precision tag for LEP2.\cite{CEEXMC}  
CEEX works at the level of spinor helicity 
amplitudes, greatly facilitating the calculation of effects such as 
ISR-FSR interference, which are included in \KKMC, and therefore in HERWIRI2.
CEEX is maximally inclusive: all real photons radiated are kept in the 
event record, no matter how soft or collinear.  There is no need to integrate
out a region of soft phase space, because the exponentiated amplitudes 
are well-behaved at $k=0$. 

The CEEX cross section for $q{\overline q}\rightarrow f{\overline f}$ has the
form \cite{KKMC}
\begin{equation}
\sigma = \frac{1}{\rm flux}\sum_{n=0}^\infty \int d{\rm PS} 
        \ \rho^{(n)}_{\rm CEEX} ({\vec p}, {\vec k}) 
\end{equation}
where
\begin{equation}
\rho^{(n)}_{\rm CEEX} = \frac{1}{n!} e^{Y({\vec p}, E_{\rm min})} 
   {1\over 4}
\sum_{{\vec\lambda},{\vec\mu}} \left|{\cal M}\left({\begin{array}{cc}
                                {\vec p} & {\vec k} \\
                                {\vec \lambda} & {\vec\mu}
                                \end{array}}\right)\right|^2 
\end{equation}
The YFS form factor is
\begin{eqnarray}
Y({\vec p}, E_{\rm min}) &=&
                          Q_i^2 Y(p_1, p_2, E_{\rm min})
                        + Q_f^2 Y(p_3, p_4, E_{\rm min})
                        + Q_i Q_f  Y(p_1, p_3, E_{\rm min})\nonumber\\
                      & &  + Q_i Q_f  Y(p_2, p_4, E_{\rm min})
                        - Q_i Q_f  Y(p_1, p_4, E_{\rm min})
                        - Q_i Q_f  Y(p_2, p_3, E_{\rm min}),
\end{eqnarray}
where $Q_i, Q_f$ are initial and final parton charges, and 
\begin{equation}
Y(p_i, p_j, E_{\rm min}) = 2\alpha{\widetilde B}(p_i,p_j, E_{\rm min}) 
                         + 2\alpha {\rm Re}\; B (p_i,p_j),
\end{equation}
with real and virtual form factors defined respectively by
\begin{eqnarray}
{\widetilde B} &=& -\int_{k^0<E_{\rm min}} \frac{d^3 k}{8\pi^2 k^0}
\left( \frac{p_i}{p_i\cdot k} - \frac{p_j}{p_j\cdot k} \right)^2, \\
B &=& \frac{i}{(2\pi)^3} \int \frac{d^4 k}{k^2}
\left(\frac{2p_i + k}{2p_i\cdot k + k^2} - \frac{2p_j - k}{2 p_j\cdot k- k^2}
\right) .
\end{eqnarray}

The $n$-photon helicity-spinor amplitude can be expanded
in terms of order $\alpha^r$ having the form 
\begin{equation}
{\cal M}_n^{(r)} = \sum_{\cal P} \prod_{i=1}^n {\cal S}_i^{({\cal P}_j)} 
\left[
\beta_0^{(r)} \left({\begin{array}{c}
                                {\vec p} \\
                                {\vec \lambda} 
                                \end{array}};~ X_{\cal P} \right) 
 + \sum_{j=1}^n 
\frac{\beta_1^{(r)} \left({\small\begin{array}{cc}
                                {\vec p} & k\\
                                {\vec \lambda} & \mu
                                \end{array}};~
X_{{\cal P}} \right)}{{\cal S}_j^{({\cal P}_j)}} 
 + \cdots + 
\sum_{1<j_1<\cdots<j_n} 
\frac{\beta_n^{(r)} \left({\small\begin{array}{cc}
                                {\vec p}& {\vec k}\\
                                {\vec\lambda} & {\vec\mu}
                                \end{array}};~
X_{{\cal P}} \right)}{{\cal S}_{j_1}^{({\cal P}_{j_1})}\cdots 
{\cal S}_{j_n}^{({\cal P}_{j_n})}} \right] 
\end{equation}
with residual spinor amplitudes $\beta_i^{(r)}$ and complex soft photon factors
${\cal S}_j$ with the property
\begin{equation}
\left|{\cal S}_{j}^{({\cal P}_{j})} \right| = -2\pi\alpha Q^2
\left( {p_a\over p_a\cdot k_j} - {p_b\over p_b\cdot k_j}\right)^2
\end{equation}
where $Q, p_a, p_b$ belong to the initial or final fermions depending on the
partition ${\cal P}_{j}$.

\KKMC\ incorporates the DIZET library (version 6.21) from the 
program ZFITTER. \cite{ZFITTER}
The $\gamma$ and $Z$ propagators are multiplied by vacuum polarization factors:
\begin{equation}
{H_\gamma} = {1\over{2-\Pi_\gamma}}, \qquad
{H_Z} = 4 \sin^2(2\theta_{\rm W}) {\rho_{\rm EW} G_\mu M_Z^2\over 8\pi \alpha 
        \sqrt{2}}.
\end{equation}
Vertex corrections are incorporated into the coupling of $Z$ to $f$ via
form factors in the vector coupling:
\begin{equation}
g_V^{(Z,f)} = {T_3^{(f)}\over \sin(2\theta_W)} - Q_f {F_{\rm v}^{(f)}(s)} \tan\theta_W.
\end{equation}
Box diagrams contain these plus a new angle-dependent form-factor in the 
doubly-vector component:
\begin{equation}
g_V^{(Z,i)} g_V^{(Z,f)} = {T_3^{(i)} T_3^{(f)} - 2T_3^{(i)} Q_f {F_{\rm v}^{(f)}(s)}
        - 2Q_i T_3^{(f)} {F_{\rm v}^{(i)}(s)} + 4Q_i Q_f {F_{\rm box}^{(i,f)} (s,t)} \over 
\sin^2(2\theta_W)} .
\end{equation}
The correction factors are calculated at the beginning of a run and stored in
tables.

The Drell-Yan cross section with multiple-photon emission
can be expressed as an integral
over the parton-level process $q_i(p_1) {\overline q}_i(p_2)\rightarrow 
f(p_3) {\overline f}(p_4) + n\gamma(k)$, integrated over phase space and
summed over photons.  The parton momenta $p_1, p_2$ are generated using
parton distribution functions giving a process at CMS
energy $q$ and momentum fractions $x_1, x_2$ such that $q^2 = x_1 x_2 s$:
\begin{equation}
\sigma_{\rm DY} = \int {dx_1\over x_1} {dx_2\over x_2}  
        \sum_i 
        f_i(q, x_1)f_{\overline i}(q, x_2)
         \sigma_i (q^2)\delta(q^2-x_1 x_2 s),
\end{equation}
where the final state phase space includes $p_3, p_4$ and $k_i$, $i=1,\cdots, n$ 
and multiple gluon radiation + hadronization is included through a shower.

The parton-level cross section $\sigma_i(q^2)$ can be calculated by \KKMC,
which integrates over a final space phase space with two fermions and an
arbitrary number of photons:
\begin{equation}
\sigma_i(q^2) =  \sum_{n=0}^\infty \int d{\rm PS}_{2+n} 
        \sigma_i ({\vec p},{\vec k})
\end{equation}

HERWIRI2 uses HERWIG 6.5 as the shower generator, which creates the 
hard process first, at Born level, in subroutine HWEPRO (HWHDYP), and then 
passes it to the cascade generator HWBGEN.
HERWIRI2 finds the $Z/\gamma^*$ and the partons interacting with it in the
event record.  The initial partons define $p_1$, $p_2$, which are transformed
to the CM frame and projected on-shell to create a starting point for \KKMC,
which generates the final fermion momenta $p_3, p_4$ and photons $k_i$ (both
ISR and FSR.) The generated particles are transformed back to the lab frame
and placed in the event record.

In addition to the basic DY process, HERWIG generates ``Compton''
events $g + q \rightarrow q + Z/\gamma^*$. About 10\% of the events have this
form. This is factorized into gluon emission times a hard EW process
at a shifted value of $q^2$. These have a different profile in the event
record, but can be processed by \KKMC\ as well.
There is also a third class of events with the emission of an
additional hard gluon. About 1\% of the events have this form, and also
have a significant shift of the $Z$ energy from its generation scale.

With a change of variables, the Drell-Yan cross section in HERWIG
can be expressed as
\begin{eqnarray}
\sigma_{\rm DY} &=& \int {dx_1\over x_1} {dx_2\over x_2}  \sum_i 
        f_i(q, x_1)f_{\overline i}(q, x_2)
         \sigma_i (q^2)\delta(q^2-x_1 x_2 s)\nonumber\\
&=& \int_{q_{\rm min}}^{q_{\rm max}} dq P(q) \int_{q^2/s}^1 {dx_1\over x_1} 
      \sum_i P_i\ W_{\rm HW}^{(i)}(q^2,x_1)
= \left\langle W_{\rm HW}\right\rangle
\end{eqnarray}
where $P(q)$ is a normalized, integrable, crude probability distribution for
$q$, $P_i$ is the crude probability of generating parton $i$, and
$W_{\rm HW}$ is the HERWIG event weight.  This weight depends only
on the hard Born cross section and is not altered by the shower.

The crude probability distributions used by HERWIG are
\begin{equation}
P(q) = \frac{1}{2}[P_\gamma(q) + P_Z(q)], \qquad 
P_\gamma(q)= {N_\gamma\over q^4}, \qquad
P_Z(q) = {N_2 q\over (q^2 - M_Z^2) + \Gamma_Z^2 M_Z^2}
\end{equation}
The HERWIG event weight is
\begin{equation}
W_{\rm HW} = \sum_i W_{\rm HW}^{(i)}, \qquad W_{\rm HW}^{(i)} = {1\over P(q)} 
        f_i(q, x_1)f_{\overline i}(q, x_2) \ln\left(s\over q^2\right) 
\sigma_{\rm HW}^{(i)}(q^2)
\end{equation}
and the corresponding probability for selecting parton $i$ is
\begin{equation}
P_i = W_{\rm HW}^{(i)}/W_{\rm HW}
\end{equation}

We have chosen to introduce electroweak corrections in a 
minimally invasive way, incorporating them in a form factor
\begin{equation}
F^{(i)}_{EW}(q^2) = {\sigma_i(q^2)\over \sigma_{\rm Born}^{(i)}(q^2)}
\end{equation}
\KKMC\ will calculate the EW form factor, and multiply it by the 
HERWIG Born cross section.
\begin{equation}
\sigma_{\rm HW+EW} = \left\langle W_{\rm Tot}\right\rangle , \qquad
W_{\rm Tot} = F_{EW}^{(i)}(q^2) W_{\rm HW} = W_{\rm HW} 
{\sigma_{\rm KK}^{(i)}(q^2) \over \sigma_{\rm Born}^{(i)}(q^2)}.
\end{equation}

The \KKMC\ cross section is calculated using a primary distribution
\begin{equation}
{d\sigma_{\rm Pri}^{(i)}(s,v) \over dv} = 
 \sigma_{\rm Born}^{(i)}(s(1-v)) 
{1\over 2}\left(1 + {1\over{\sqrt{1-v}}}\right)
{\overline\gamma}_i v^{{\overline\gamma}_i-1} v_{\rm min}^{\gamma_i -
                        {\overline\gamma}_i}
\end{equation}
with
\begin{equation}
\gamma_i = {2\alpha\over\pi}Q_i^2 
\left[\ln \left({s\over m_i^2}\right)  - 1\right],
\qquad {\overline\gamma}_i 
        = {2\alpha\over\pi}Q_i^2\ln\left({s\over m_i^2}\right)
\end{equation}
to generate the factor $v$ giving the fraction of $s$ remaining after ISR photon
emission, $s_X = s(1 - v)$.

The \KKMC\ cross section is
\begin{equation}
\sigma(q^2) = 
\int d\sigma_{\rm Pri} {d\sigma_{\rm Cru} \over d\sigma_{\rm Pri}}
{d\sigma_{\rm Mod}\over d\sigma_{\rm Cru}} 
= \sigma_{\rm Pri} \left\langle W_{\rm Cru} W_{\rm Mod}
\right\rangle .
\end{equation}
$W_{\rm Cru}$ is calculated during ISR generation and
$W_{\rm Mod}$ is generated after $s_X$ is available.

The HERWIG and \KKMC\ weights are combined to calculate the total 
HERWIRI2 weight, 
\begin{equation}
\sigma_{\rm Tot} = \left\langle W_{\rm HW} 
        {\sigma_i(q^2)\over \sigma_{\rm Born}^{(i) \star}(q^2)}\right\rangle 
= \left\langle W_{\rm HW} \sigma_{\rm Pri}^{(i)}(q^2) 
        {W_{\rm Cru}^{(i)} W_{\rm Mod}^{(i)} 
        \over \sigma_{\rm Born}^{(i) \star}(q^2) }
\right \rangle, 
\end{equation}

This average {\it could} be calculated using
a joint probability distribution for $q$ and $v$,
$D(q,v) = P(q) d\sigma_{\rm Pri}/dv$, with $P(q)$ from HERWIG.
An adaptive MC (S. Jadach's FOAM \cite{FOAM}) could calculate
the normalization of the distribution at the beginning of the run, in a
similar manner to how \KKMC\ presently integrates the one-dimensional
primary distribution.  To account for beamsstrahlung, \KKMC\ already permits
such a user-defined distribution, in up to three variables.
However, as a first step, we have tried to run HERWIRI2 using 
\KKMC's one-dimensional primary distribution. This requires fixing an overall 
scale $q_0$ to initialize \KKMC\ ({\it e.g.}, $q_0 = M_Z$).

The built-in primary distribution for electrons at scale $q_0$ can be used
for the low-level generation of $v$.
The transformation from this distribution to a distribution at HERWIG's
generated scale $q$ for quark $i$ is then obtained by a change of variables:
\begin{equation}
\sigma_{\rm Tot} = \sigma_{\rm Pri}^{(e)} 
       \left\langle W_{\rm HW}\left(
 {d\sigma_{\rm Pri}^{(i)}(q^2,v)\over
                     d\sigma_{\rm Pri}^{(e)}(q_0^2,v)}
   \right) \left(
        {W^{(i)}_{\rm Crud} W^{(i)}_{\rm Mod}\over 
\sigma_{\rm Born}^{(i) \star}(q^2)}\right)
\right\rangle
\end{equation}
with
\begin{equation}
\frac{d\sigma_{\rm Pri}^{(i)}(q^2,v)}{
                     d\sigma_{\rm Pri}^{(e)}(q_0^2,v)}
                =  W_{\gamma}^{(i)}\frac{\sigma_{\rm Born}^{(i)}(q^2(1-v))}{
                                      \sigma_{\rm Born}^{(e)}(q_0^2(1-v))} \ ,
\end{equation}
where
\begin{equation}
W_\gamma = \frac{{\overline\gamma}_i}{{\overline\gamma}_e}
        \left(\frac{v}{v_{\rm min}}\right)^{{\overline\gamma_i} 
	- {\overline\gamma_e}} v_{\rm min}^{\gamma_i - \gamma_e} .
\end{equation}
The $\gamma$ factors are calculated using $q^2/m_i^2$ for parton $i$ and
$q_0^2/m_e^2$ for the electron.

Shuffling the numerators and denominators about gives the expression used
in HERWIRI2:
\begin{equation}
\sigma_{\rm Tot} = \left\langle W_{\rm HW} W_{\rm Mod} W_{\rm Karl} W_{\rm FF}
W_{\gamma} \right\rangle
\end{equation}
with two new weights
\begin{equation}
W_{\rm Karl} = \frac{\sigma_{\rm Pri}^{(e)} W_{\rm Crud}^{(i)}}{
                \sigma_{\rm Born}^{(e)}(q_0^2(1-v))} ,\qquad
W_{\rm FF} = \frac{\sigma_{\rm Born}^{(i)}(q^2(1-v))}{ 
                \sigma_{\rm Born}^{(i) \star} (q^2)}.
\end{equation}

HERWIRI2 is still under development, so any numerical results must be treated 
as preliminary.  A run for $pp$ collisions at 10 TeV with the
$Z/\gamma^*$ invariant mass bounded by 40 GeV and 140 GeV, using HERWIG 6.520 
default parameters and PDFs, yields a cross-section of $1183.7\pm 1.3$ pb,
compared to $1098.8\pm 1.0$ pb for HERWIG alone, giving an electroweak 
contribution of 7.7\%. Turning on ISR gives 
a much wider weight distribution and consequently, greatly reduced efficiency.  
Preliminary results give a cross-section of $1212\pm 109$ pb, 
showing an additional 2.4\% contribution from ISR.

Work is in progress to
optimize MC generation in the presence of ISR.
Once HERWIRI2 is complete and thoroughly tested, it will be 
compared to other available hadronic/electroweak generators. In particular, it
will be interesting to see the effect of initial state radiation, which is not
present in the other programs, but appears to enter at the 2 -- 3\% level, 
making it crucial to precision calculations. 

\section*{Acknowledgments}
S. Yost thanks the organizers of RADCOR 2011 for the invitation to present these
results, and D. Marlow and the Princeton Physics Department for their 
support and hospitality during a critical period of its development. S. Yost 
and B.F.L. Ward also acknowledge the hospitality of the CERN theory division,
which contributed greatly to the progress of this work.


\begin{thebibliography}{99}
\bibitem{EWsys}{N.E. Adam, V. Halyo, and S.A. Yost, {\sl JHEP}\ {\bf 05} 
(2008) 062 [arXiv:0802.3251]; {\it ibid.}, {\sl JHEP}\ {\bf 11} (2010) 074 
[arXiv:1006.3766]; N.E. Adam, V. Halyo, S.A. Yost, and W.-H. Zhu, 
{\sl JHEP}\ {\bf 09} (2008) 133 [arXiv:0808.0758].}
\bibitem{horace}{C.M. Carloni Calame, G. Montagna, O. Nicrosini, and 
M. Treccani, {\sl JHEP} {\bf 05} (2005) 019 [arXiv:hep-ph/0502218]; 
C.M. Carloni Calame, G. Montagna, O. Nicrosini, and A. Vicini,
{\sl JHEP} {\bf 12} (2006) 016 [arXiv:hep-ph/0609170]; {\it ibid.},
{\sl JHEP} {\bf 10} (2007) 109 [arXiv:0710.1722].}
\bibitem{photos}{E. Barberio, B. van Eijk, and Z. W{\c a}s, 
{\sl Comput.\ Phys.\ Commun.} {\bf 66} (1991) 115;
E. Barberio and Z. W{\c a}s, {\sl Comput.\ Phys.\ Commun.} {\bf 79} (1994) 291;
P. Golonka and Z. W{\c a}s, {\sl Eur.\ Phys.\ J.} {\bf C45}
(2006) 97 [arXiv:hep-ph/0506026].}
\bibitem{winhac}{W. P{\l}aczek and S. Jadach, {\sl Eur.\ Phys.\ J.} {\bf C29}
(2003) 325 [arXiv:hep-ph/0302065];
W. P{\l}aczek, PoS(EPS-HEP2009) 340 [arXiv:0911.0572].}
\bibitem{zinhac}{A. Si{\'o}dmok and W. P{\l}aczek, 
http://th-www.if.uj.edu.pl/ZINHAC/ .}
\bibitem{MRSTQED}{A.D. Martin, R.G. Roberts, W.J. Stirling, and R.S. Thorne, 
{\sl Eur.\ Phys.\ J.} {\bf C39} (2005) 155.}
\bibitem{KKMC}{S. Jadach, B.F.L.\ Ward, and Z. W{\c a}s, 
{\sl Comput.\ Phys.\ Commun.} {\bf 130} (2000) 130.}
\bibitem{YFS}{D.R. Yennie, S. Frautschi, and H. Suura, {\sl Ann.\ Phys.}
{\bf 13} (1961) 379.}
\bibitem{DIZET}{A. Akhundov, D. Bardin, M. Bilenky, P. Christova, S. Riemann,
T. Riemann, M. Sachwitz, and H. Vogt, DIZET6.21}
\bibitem{ZFITTER}{A. Akhundov, D. Bardin, and T. Riemann, {\sl Phys.\ Lett.}
{\bf B166} (1986) 111.}
\bibitem{QCDxQED}{C. Glosser, S. Jadach, B.F.L.\ Ward, and S.A.\ Yost, 
{\sl Mod.\ Phys.\ Lett.} {\bf A19} (2004) 2113 [arXiv:hep-ph/0404087]; 
B.F.L.\ Ward, C. Glosser, S. Jadach, and S.A.\ Yost, {\sl Int. J. Mod.\ Phys.} 
{\bf A20} (2005) 3735 [arXiv:hep-ph/0411047]; 
B.F.L.\ Ward and S.A.\ Yost, in {\sl Proc. ICHEP04, Beijing}, vol. 1, 
ed. H. Chen {\it et al.} (World Scientific, Singapore, 2005), 588 
[arXiv:hep-ph/0410277]; 
{\it ibid.}, {\sl Acta Phys.\ Polon.} {\bf B38} (2007) 2395 [arXiv:0704.0294];
{\it ibid.}, in {\sl Proc. ICHEP06, Moscow}, vol. 1, 505 [arXiv:hep-ph/0610230];
{\it ibid.}, PoS (RAD COR 2007) 038 [arXiv:0802.0724]; 
{\it ibid.}; B.F.L.\ Ward, S.\ Joseph, S.\ Majhi, and S.A.\ Yost, 
{\sl Proc. 2008 HERA-LHC Workshop}, DESY-PROC-2009-02, 
eds. H. Jung, A. De Roeck (DESY, Hamburg, 2009) 180 [arXiv:0808.3133]; 
{\it ibid.}, in {\sl Proc. ICHEP08, Philadelphia} [arXiv:0810.0723].}
\bibitem{HERWIG}{G. Corcella, I.G. Knowles, G. Marchesini, S. Moretti, 
K. Odagiri, P. Richardson, M.H.\ Seymour, and B.R.\ Webber, HERWIG6.5, 
arXiv:hep-ph/0011363.}
\bibitem{HERWIRI1}{S. Joseph, S. Majhi, B.F.L.\ Ward and S.A.\ Yost, 
{\sl Phys. Lett.} {\bf B685} (2010) 283 [arXiv:0906.0788]; 
{\it ibid.}, {\sl Phys.\ Rev.} {\bf D81} (2010) 076008 [arXiv:1001.1434]; 
{\it ibid.}, {\sl Mod.\ Phys.\ Lett.} {\bf A25} (2010) 2207; 
in {\sl Proc. DPF-2009, Detroit} eConf C090726 [arXiv:0910.0491].}
\bibitem{kernels}{B.F.L.\ Ward, {\sl Adv.\ High Energy Phys.} {\bf 2008}
(2008) 682312; {\sl Ann.\ Phys.} 323 (2008) 2147.}
\bibitem{HERWIRI1NLO}{S. Joseph, S. Majhi, B.F.L.\ Ward and S.A.\ Yost,
 Pos (RADCOR2009) 070 [arXiv:1001.2730].}
\bibitem{MCNLO}{S. Frixione and B. Webber, {\sl JHEP} {\bf 0206} (2002) 029;
S. Frixione, P. Nason, and B. Webber, {\sl JHEP} {\bf 0308} (2003) 007.}
\bibitem{POWHEG}{S. Alioli, P. Nason, C. Oleari, and E. Re, {\sl JHEP}
{\bf 0807} (2008) 060 [arXiv:0805.4802]; {\it ibid.}, {\sl JHEP} {\bf 1101}
(2011) 095 [arXiv:1009.5594].}
\bibitem{QTS3-RADCOR05}{S.A.\ Yost, Chris Glosser, and B.F.L.\ Ward, in 
{\sl Third International Symposium on Quantum Theory and Symmetries}, 
Cincinnati, 2003 (World Scientific, Singapore, 2004) 775 [arXiv:hep-ph/0401211];
S.A.\ Yost and B.F.L.\ Ward, {\sl Nucl.\ Phys.\ (Proc.\ Supp.)} 
{\bf B157} (2006) 78 [arXiv:hep-ph/0602030].}
\bibitem{bhlumi}{S. Jadach, E. Richter-W{\,a}s, B.F.L. Ward, and Z. W{\c a}s, 
{\sl Comput.\ Phys.\ Commun.} {\bf 70} (1992) 305;
S. Jadach, W. P{\l}aczek, E. Richter-W{\,a}s, B.F.L.\ Ward,
and Z. W{\c a}s, {\sl Comput.\ Phys.\ Commun.} {\bf 102} (1997) 229;
B.F.L.\ Ward, S. Jadach, M. Melles, and S.A. Yost, {\sl Phys.\ Lett.}
{\bf B450} (1999) 262.}
\bibitem{bhwide}{S. Jadach, W. P{\l}aczek, and B.F.L.\ Ward, {\sl Phys.\ Lett.}
{\bf B390} (1997) 298.}
\bibitem{koralz}{S. Jadach, B.F.L.\ Ward, and Z. W{\c a}s, 
{\sl Comput.\ Phys.\ Commun.} {\bf 79} (1994) 503.}
\bibitem{koralw}{M. Skrzypek, S. Jadach, W. P{\l}aczek, and Z. W{\c a}s, 
{\sl Comput.\ Phys.\ Commun.} {\bf 94} (1996) 216;
S. Jadach, W. P{\l}aczek, M. Skrzypek, B.F.L. Ward, and 
Z. W{\c a}s, {\sl Comput.\ Phys.\ Commun.} {\bf 119} (1999) 272.}
\bibitem{yfsww3}{S. Jadach, W. P{\l}aczek, M. Skryzpek, B.F.L. Ward, and 
Z. W{\c a}s, {\sl Comput.\ Phys.\ Commun.} {\bf 140} (2001) 432.}
\bibitem{ZF1}{D. Bardin, P. Christova, and O. Fedorenko, {\sl Nucl.\ Phys.}
{\bf B175} (1980) 435; {\sl Nucl.\ Phys.} {\bf B197} (1982) 1;
A. Akhundov, D. Bardin, and T. Riemann, {\sl Nucl.\ Phys.}
{\bf B276} (1986) 1;
D. Bardin, S. Riemann, and T. Riemann, {\sl Z. Physik} {\bf C32} (1986) 121;
D. Bardin, S. Bilenky, G. Mitselmakher, T. Riemann, and M.  Sachwitz, 
{\sl Z. Physik} {\bf C44} (1989) 493;
D. Bardin, W. Hollik, and T. Riemann, {\sl Z. Physik} {\bf C49} (1991) 485.
}
\bibitem{ZF2}{
A. Djouadi and C. Verzegnassi, {\sl Phys.\ Lett.} {\bf B195} (1987) 265; 
A. Djouadi, {\sl Nuovo Cim.} {\bf 100A} (1988) 357.}
\bibitem{ZF3}{H. Burkhardt, F. Jegerlehner, G. Pensko, and C. Verzegnassi, 
{\sl Z. Physik} {\bf C43} (1989) 497;
F. Jegerlehner, in {\sl Progress in Particle and Nuclear Physics},
 Vol. 27, ed. A. Fassler (Pergamon Press, Oxford, 1991) 32.}
\bibitem{CEEXMC}{S. Jadach, B.F.L. Ward, and Z. W{\c a}s, {\sl Phys.\ Rev.}
{\bf D63} (2001) 113009 [arXiv:hep-ph/0006359].}
\bibitem{FOAM}{S. Jadach, {\sl Comput.\ Phys.\ Commun.} {\bf 130} (2000) 244
[arXiv:physics/99100004].}
\end{thebibliography}
\end{document}